\documentclass[aps,twocolumn,amsmath,amssymb,physics,array,orcidlink,MnSymbol]{revtex4-2}
\usepackage[colorlinks=true,linktoc=page,linkcolor=blue,citecolor=blue,urlcolor=blue]{hyperref}
\usepackage{physics,array,orcidlink,MnSymbol,float,marvosym,flushend}
\definecolor{cyan}{HTML}{00FFFF}
\newcommand{\la}{{\langle}}
\newcommand{\ra}{{\rangle}}
\allowdisplaybreaks
\begin{document}

\title{Kitaev model in a magnetic field: stable emergent structure, degenerate classical ground states, and reentrant topology}	

\author{Sheikh Moonsun Pervez~\orcidlink{0009-0000-2283-053X}}
\email[\Letter~]{moonsun@iopb.res.in}
\affiliation{Institute of Physics, Sachivalaya Marg, Bhubaneswar-751005, India}
\affiliation{Homi Bhabha National Institute, Training School Complex, Anushakti Nagar, Mumbai 400094, India}

%******************************************************************************%
%********************************** ABSTRACT **********************************%
%******************************************************************************%

\begin{abstract}
We have studied the anti-ferromagnetic Kitaev model on a honeycomb lattice under the Zeeman field, using an extensive Majorana mean-field analysis. When the magnetic field is along a specific Cartesian axis, we find that the emergent fields exhibit direction-dependent stabilization up to a certain critical strength of the external field. For a conical magnetic field, the characteristics of the emergent intermediate state are elusive. Our mean-field analysis reveals the existence of two distinct phases in the intermediate region. First, the system enters a disordered phase, where emergent-field densities converge to random values, and the Chern number is ill-defined. The magnitude of magnetization also fluctuates and remains less than unity, indicating a strong quantum effect. In the second phase, emergent-field densities attain vanishingly small values. In this phase, the magnetization components fluctuate heavily, but the magnitude of the magnetization vectors becomes unity, indicating highly degenerate classical ground states. We perform exact diagonalization calculations that qualitatively support some of the mean-field results. We extend our study to the anisotropic limit of the Kitaev coupling parameters. When the couplings are beyond the triangular inequality, the pure Kitaev model is known to host a topologically trivial gapped quantum spin liquid. We find that, for intermediate strengths of a conical magnetic field, topology shows a reentrant behavior.
\end{abstract}

\date{\today}
\maketitle

%******************************************************************************%
%************************ SECTION - 1: INTRODUCTION ***************************%
%******************************************************************************%

\section{Introduction}\label{section_introduction}
Highly entangled quantum phases, such as spin liquids, are exhibited by frustrated magnetic systems~\cite{review_balents_nature,review_balents}. Kitaev proposed such a system with direction-dependent Ising interactions on a honeycomb lattice~\cite{kitaev_2006}, which shows the signature of quantum spin liquid (QSL). Depending on the relative strength among the different bond-dependent exchange couplings, this QSL can be gapless or gapped. Besides being an exactly solvable interacting system in two~\cite{saptarshi_2007,mandaljpa} and three dimensions~\cite{naveen_2008}, this model has potential application to topological quantum computation~\cite{kitaev_2006}. A large number of works soon flooded in~\cite{review_lee,review_knolle,takagi_2019,review_trebst} after Jackeli and Khaliullin proposed a mechanism to realize the Kitaev model in Mott insulators~\cite{jackeli_khaliullin}. Lots of experimental works have been done on the Kitaev candidate materials $\alpha$-RuCl$_3$, Na$_2$IrO$_3$, $\beta$-Li$_2$IrO$_3$, etc.~\cite{abanerjee_science,a2iro3,na2iro3,li2iro3,a2iro3_HK_model}, but the search is still ongoing. The main difficulty is the presence of non-Kitaev interactions (Heisenberg, $\Gamma$, $\Gamma'$, etc.) in real materials~\cite{trebst_review_kitaev_materials}. The Kitaev candidate materials always exhibit some magnetic order at very low temperatures, which makes them not suitable to realize the QSL~\cite{primer_on_kitaev_model_saptarshi}. However, studies have been done on extended-Kitaev models to realize proximate QSL~\cite{abanerjee_2016,proximate_ksl_K_J_Gamma,kitaev_quasiparticle_in_proximate_QSL}.
\begin{figure}[h]\centering 
\includegraphics[width=0.94\columnwidth,height=!]{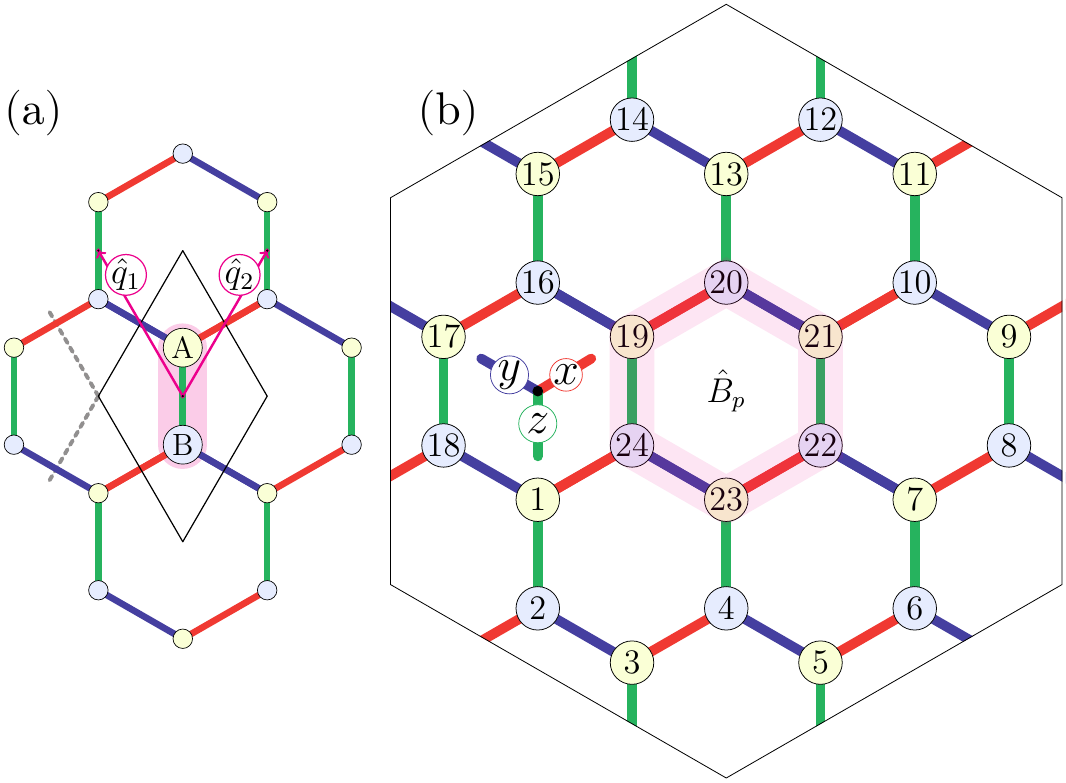}
\caption{(a) Representation of the momentum space vectors (magenta arrows) $\hat{q}_1$ and $\hat{q}_2$ in the rhombic Brillouin zone (in black line). The two-site unit cell is shaded in magenta. The dashed gray lines are shown to easily identify the correspondence between the hexagon and rhombus. (b) 24-site Kitaev honeycomb model with periodic boundary conditions used for the exact diagonalization. The direction-dependent Ising-like $x$, $y$, $z$-type interactions are indicated in red, blue, and green colours. One of the plaquette operators $\hat{B}_p$, is shown in magenta shade.}\label{figure_schematics}
\end{figure}
\\\\\indent
To study the effect of magnetic field on Kitaev spin liquid (KSL), Tikhonov {\it et al.} performed a perturbative analysis that shows the nearest-neighbor spin-spin correlation builds up quadratically with the magnitude of the external perturbation~\cite{Kitaev_2011}. Later, the model was numerically investigated in the non-perturbative region too~\cite{KSL_in_magnetic_fields_dynamical_response,kitaev_gamma_zeeman_phase_diagram,emergent_glassiness_yogendra,emergent_magnetic_order_KSL_111,dynamical_topological_properties_KSL_111,robust_non_abelian_gapless_prb,fractionalization_signature_KSL_111,tuning_topological_order_PRL,kitaev_model_theory_in_111_natcomm,u1_KSL_natcom,magnetic_field_driven_qpt_nandini,two_magnon_bound_state,naturetopologicaltransition111,majorana_metal_nandini}. All studies suggest that QSL survives up to a critical field before entering into an intermediate phase, whose exact characteristics remains elusive for a conical magnetic field. It has been suggested to be gapless~\cite{dynamical_topological_properties_KSL_111,robust_non_abelian_gapless_prb,fractionalization_signature_KSL_111}, gapped~\cite{tuning_topological_order_PRL,kitaev_model_theory_in_111_natcomm}, $U(1)$ gapless~\cite{u1_KSL_natcom,magnetic_field_driven_qpt_nandini,two_magnon_bound_state}, chiral spin liquid~\cite{naturetopologicaltransition111}, or Majorana metal phase~\cite{majorana_metal_nandini}. This motivates us to investigate the Kitaev model in a magnetic field once more.
\\\\\indent
In this paper, we investigate the anti-ferromagnetic Kitaev model in an external magnetic field directed along $[0,0,1]$ and $[1,1,1]$ directions. We have executed an extensive mean-field (MF) analysis with a large number of {\it independent} parameters to hunt for the nature of this intermediate phase, followed by exact diagonalization (ED) calculations to support the MF analysis. We extend the study with $[1,1,1]$ magnetic field to the anisotropic region of the Kitaev coupling parameters.
\\\\\indent
We have found, magnetic field along the $\hat{z}$-direction immediately destabilizes the gauge field defined on the $z$-type bond. But those defined on $x$ or $y$-type bonds are stable up to two critical fields. This is in partial agreement with the results obtained in our previous work with small Kitaev clusters, where the static/ dynamic nature of emergent fields has been explicitly shown in time~\cite{kitaev_cluster_dynamical_pervez}. Stabilization of these fields indicates partially robust QSL~\cite{shifeng_anyon_dynamics,feng2026magneticfieldinducedphenomena,kitaev_cluster_static_pervez}, and it is topologically non-trivial with Chern number $C_n=+1$. Beyond the first critical field, $C_n=-1$. After the second critical field, the gauge fields defined on $x$ and $y$ bonds also become dynamic, indicating the commencement of another QSL, which is topologically trivial. The emerging structure is completely destroyed beyond a third critical field, identifying the spin-polarized (SP) phase.
\\\\\indent
The situation is particularly interesting for a conical magnetic field. All of the emergent gauge fields are equally dynamic, in a topologically non-trivial phase with $C_n=-1$. Beyond a first critical field, the system enters into an intermediate phase, where the emergent fields' densities are arbitrary, magnetization components fluctuate, and the Chern number is ill-defined (fluctuates between $C_n=+2$~\cite{tuning_topological_order_PRL,kitaev_gamma_zeeman_phase_diagram,naturetopologicaltransition111} and $C_n=0$), which is in accordance with MF results from a previous study~\cite{emergent_magnetic_order_KSL_111}. Beyond a second critical field, the system becomes topologically trivial, where the Cartesian components of magnetization fluctuate heavily, though they obey the constraint of $|\vec{m}_{^{A/B}}|=1$ ($A/B$ are the two sub-lattice indices of the honeycomb lattice). In this phase, the emergent structures become vanishingly small. Studies in existing literature do not treat all of the magnetization components as separate MF parameters. That is why this particular structure remained hidden in all previous MF studies. We attribute this feature as a signature of large degeneracy in classical ground states (as the magnetization operator's expectation behaves like a classical vector). Thus, regarding the large degeneracy of ground states, we have partial agreement with the previous works with ED, which showed a plethora of states come down in the intermediate state~\cite{u1_KSL_natcom,magnetic_field_driven_qpt_nandini,robust_non_abelian_gapless_prb}. It is not a spin liquid under the present MF analysis, as the emergent structures are almost destroyed in this state. At a third critical field, the system enters the SP phase. Our ED calculations qualitatively support some of the MF results.
\\\\\indent
Finally, we consider anisotropy in the exchange couplings and subject the Hamiltonian to a conical magnetic field. The gapless phase of the pure Kitaev model immediately becomes topologically non-trivial with $C_n=-1$. Interestingly, at an anisotropic parameter point beyond the triangular inequality, the gapped topologically trivial QSL of the pure Kitaev model shows a transition into a topologically non-trivial phase for intermediate strengths of the magnetic field. We explain the reentrance of topology in terms of some effective couplings entering the triangular inequality.
\\\\\indent
The paper is structured as follows. In section~\ref{section_model_and_method}, we introduce the model under consideration and provide details of the MF decomposition. In section~\ref{section_results}, we tabulate the MF results in~\ref{section_MFresults}, followed by ED support in~\ref{section_EDresults}. The reentrant topology in the anisotropic Kitaev model is investigated in~\ref{section_anisotropy_results}. We conclude in section~\ref{section_conclusion}.

%******************************************************************************%
%****************** SECTION - 2 : MODEL AND METHOD ****************************%
%******************************************************************************%

\section{Model and Method}\label{section_model_and_method}
\subsection{Hamiltonian}\label{section_model}
Kitaev model on honeycomb lattice is a spin 1/2 system where the nearest neighbor spin-spin interaction is direction-dependent Ising-type. The system Hamiltonian is given by~\cite{kitaev_2006},
\begin{eqnarray}
H_{\rm K}&=&\kappa_x\sum_{\la j,k\ra_x}\sigma_j^{x}\sigma_k^{x}+\kappa_y\sum_{\la j,k\ra_y}\sigma_j^{y}\sigma_k^{y}+\kappa_z\sum_{\la j,k\ra_z}\sigma_j^{z}\sigma_k^{z},\label{equation_kitaev_model}
\end{eqnarray}
where $\vec{\sigma}_j$ is the Pauli spin 1/2 operator at site $j$, and $\la j,k\ra_\alpha$ is the $\alpha$-type bond that connects nearest neighbors $j$ and $k$. While studying the isotropic case, we have considered $\kappa_x=\kappa_y=\kappa_z=\kappa$, and adopt the natural units by letting $\hbar=1$. All of the results are obtained by normalizing the Hamiltonian by the Kitaev coupling, i.e., in units of $|\kappa|$. After splitting the spins into Majorana fermions, this model can be written as a four-body Majorana hopping model. These Majoranas can be re-coupled to form gauge fields which make the background static $Z_2$ structure, on which, the emergent `matter fermions' hop. In a particular choice of the gauge sector, the Hamiltonian reduces to a two-body hopping model for the matter-fermions, which can be solved exactly, and the result is KSL~\cite{kitaev_2006}. A spin operator $\sigma_j^\alpha$ when operates on an eigen-state $\ket{\psi}$, it flips the gauge sector by injecting $\pi$ flux on the two plaquettes adjacent to the $\alpha$ bond originated from $j^{\rm th}$ site, apart from introducing an itinerant Majorana to the matter sector. This change can be non
vanishingly compensated by $\sigma_k^\beta$ , only if $\beta=\alpha$, and $|\vec{r}_j-\vec{r}_k|$ = 0 (trivially, following Pauli algebra)
or 1 (actual flux compensation). This results in a short range correlation which cuts off beyond the nearest neighbor, indicating an extreme liquid nature of the eigen-state~\cite{saptarshi_2007}. In presence of an external field, the exact solvability is destroyed, and the problem has to be tackled numerically, for large field values. When subjected to an external magnetic field $\vec{h}$, the Kitaev-Zeeman (KZ) Hamiltonian becomes,
\begin{eqnarray}
H_{\rm KZ}&=&H_{\rm K}-\vec{h}\cdotp\sum_{j=1}^{N}\vec{\sigma}_j,\label{equation_kitaev_zeeman_model}
\end{eqnarray}
where $N$ is the total number of sites. A Kitaev system with periodic boundary condition for $N=24$~\cite{kitaev_heisenberg_khaluillin} is depicted in FIG.\ref{figure_schematics}(b), which has been used to do the ED calculations. Two magnetic field directions under consideration are $\hat{h}_{001}=\hat{z}$, and $\hat{h}_{111}=(\hat{x}+\hat{y}+\hat{z})/\sqrt{3}$. With this set-up, we are now ready to delve into the MF decomposition that we have performed in this work.
\subsection{Mean-field decomposition}\label{section_method}
\subsubsection{Fermionization and the mean fields}
As we are concentrating on KZ system, all the interaction terms are of the type $\sigma_j^\alpha\sigma_k^\beta$, with $\beta=\alpha$ only, and this interaction is defined on a $\alpha$-type bond. Using Kitaev fermionization process, we can split the spin at a site into four Majoranas (three of them are indexed `$b$' Majoranas, and one of them is non-indexed `$c$' Majorana)~\cite{kitaev_2006}. We always consider $j$-site belongs to `$A$' sub-lattice and $k$-site belongs to `$B$' sub-lattice. We put this sub-lattice index too in the Majorana definition, so that when we go to Fourier space and the nearest neighbor interacting terms are written using the same site index along with a translation vector (something like $\vec{r}_k=\vec{r}_j+\vec{v}$), we don't loose the track of which Majorana belongs to what sub-lattice. So, on sub-lattice $A$ and $B$, spins are split as,
\begin{eqnarray}
\sigma_{j,A}^{\alpha}=i~b_{j,A}^{\alpha}~c_{j,A}~,\qquad	\sigma_{k,B}^{\alpha}=i~b_{k,B}^{\alpha}~c_{k,B}.\label{equation_kitaev_fermioniaztion}
\end{eqnarray}
Under MF decomposition, each interacting term can be approximated as,
\begin{eqnarray}
\sigma_j^{\alpha}\sigma_k^{\alpha}&\approx&m_A^{\alpha}~(ib_{k,B}^{\alpha}c_{k,B})+(ib_{j,A}^{\alpha}c_{j,A})~m_B^{\alpha}-m_A^{\alpha}m_B^{\alpha} \nonumber\\
&-&g^{\alpha}_{\gamma}~(ic_{j,A}c_{k,B})-(ib_{j,A}^{\alpha}b_{k,B}^{\alpha})~f_{\gamma}+g^{\alpha}_{\gamma}f_{\gamma}\nonumber\\
&-&u^{\alpha}_{\gamma}~(ib_{k,B}^{\alpha}c_{j,A})-(ib_{j,A}^{\alpha}c_{k,B})~w^{\alpha}_{\gamma}+u^{\alpha}_{\gamma}~w^{\alpha}_{\gamma}~,\label{equation_mean_field_decomposition}
\end{eqnarray}
where the nearest neighbors $j,k$ are connected by a $\gamma$-type bond, and in general $\gamma$ may not be equal to $\alpha$. The defined MF variables are, 
\begin{eqnarray}
\la ib_{j,A}^{\alpha}c_{j,A}\ra=m_A^{\alpha}~,\qquad\la ib_{k,B}^{\alpha}c_{k,B}\ra=m_B^{\alpha}.\label{equation_mf_variable_magnetization}
\end{eqnarray}
These are components of magnetization on each sub-lattice, so there are total six of them. Next mean fields are gauge and matter sector correlations,
\begin{eqnarray}
\la ib_{j,A}^{\alpha}b_{k,B}^{\alpha}\ra_\gamma=g^{\alpha}_{\gamma},\qquad\la ic_{j,A}c_{k,B}\ra_\gamma=f_{\gamma}~.\label{equation_mf_variable_emergent_fields}
\end{eqnarray}
We have to take either $g^\alpha_\gamma=g^\alpha~\forall\gamma$, or $g^\alpha_\gamma=g_\gamma~\forall\alpha$, as these two scenarios are numerically indistinguishable; we consider the later choice to keep it consistent with matter field definitions. So there are three of the gauge fields and three matter fields. The next fields can be thought of as gauge-matter transition fields,
\begin{eqnarray}
\la ib_{j,A}^{\alpha}c_{k,B}\ra_\gamma=u^{\alpha}_{\gamma}~,\qquad
\la ib_{k,B}^{\alpha}c_{j,A}\ra_\gamma=w^{\alpha}_{\gamma}.\label{equation_mf_variable_gauge_matter_transition}
\end{eqnarray}
We have taken $u^\alpha_\gamma=u$ and $w^\alpha_\gamma=w$, for all $\alpha,\gamma=x,y,z$. These careful choices of the MF variables come after trying out many different MF ansätze. Because of the presence of large number of MF variables, to reliably converge the MF, we started with zeroth order MF decomposition (simplest choices of the MF variables: $m_A^\alpha=m_A$, $m_B^\alpha=m_B$, $u_{\alpha}=u$, $w_{\alpha}=w$, $f_{\alpha}=f$, $g_{\alpha}=g$, for all $\alpha=x,y,z$), and gradually increased the complexity in the analysis. At each order, we observe specific pattern among the variables, and in next order, we implement that (for example, if two {\it independent} variables from the same MF species are converged to similar value for the full range of $h$, we treat them as a single variable in the next order), while incorporating more number of MF variables. Results from different MF ansätze are consistent.
\subsubsection{Lagrange multiplier}
Kitaev fermionizaion locally enlarges the Hilbert space by a factor of two, which needs to be projected. The projection operator at site $j$ is $D_j=b_j^x b_j^y b_j^z c_j=(-)1$ for (un)physical subspace. The constraint is difficult to implement in this quartic form, but we can transform it to a set of quadratic equations as: $b_j^x b_j^y+b_j^z c_j=0$ (and other cyclic permutations). Introducing the Lagrange multipliers~\cite{tuning_topological_order_PRL}, the constrained Hamiltonian becomes,
\begin{eqnarray}
H_c&=&H_{\rm KZ}-\sum_{j\in A}\sum_{\alpha=x,y,z}\lambda_A^\alpha\left(ib_{j,A}^\alpha c_{j,A}+\frac{i}{2}\varepsilon^{\alpha\beta\gamma} b_{j,A}^\beta b_{j,A}^\gamma\right)\nonumber\\
&&-\sum_{k\in B}\sum_{\alpha=x,y,z}\lambda_B^\alpha\left(ib_{k,B}^\alpha c_{k,B}+\frac{i}{2}\varepsilon^{\alpha\beta\gamma} b_{k,B}^\beta b_{k,B}^\gamma\right),\label{equation_constrained_hamiltonian}
\end{eqnarray}
where $\varepsilon^{\alpha\beta\gamma}$ is the purely anti-symmetric Levi-Civita symbol. After trying out different combinations for $\lambda$'s, we are convinced that a site and direction-independent single Lagrange multiplier is enough to provide good convergence and consistent results.
\subsubsection{Momentum space Hamiltonian}
Next we do the Fourier transformation of each of the Majoranas to go to momentum $\vec{q}$ space,
\begin{eqnarray}
b_{j(k),A(B)}^{\alpha}=\frac{1}{\sqrt{N}}\sum_{\vec{q}}
e^{i\vec{r}_{j(k)}\cdotp\vec{q}}~b_{\vec{q},A(B)}^{\alpha},\nonumber\\
c_{j(k),A(B)}=\frac{1}{\sqrt{N}}\sum_{\vec{q}}
e^{i\vec{r}_{j(k)}\cdotp\vec{q}}~c_{\vec{q},A(B)},\label{equation_fourier_transformation}
\end{eqnarray}
with the normalization, $\frac{1}{N}\sum_{j}e^{i\vec{r}_j\cdotp\left(\vec{q}-\vec{q'}\right)}=\delta(\vec{q}-\vec{q'})$. After MF decomposition, the quartic Hamiltonian becomes quadratic, and it is written in the basis of $\bra{\Psi}=\left(b^{x^\dagger}_{\vec{q},A}\quad b^{y^\dagger}_{\vec{q},A}\quad b^{z^\dagger}_{\vec{q},A}\quad c^{\dagger}_{\vec{q},A}\quad b^{x^\dagger}_{\vec{q},B}\quad b^{y^\dagger}_{\vec{q},B}\quad b^{z^\dagger}_{\vec{q},B}\quad c^{\dagger}_{\vec{q},B}\right)$, as:
\begin{eqnarray}
H_{\rm KZ}=\sum_{\vec{q}\in {\rm HBZ}}\bra{\Psi} H_{\vec{q}} \ket{\Psi}+\Sigma.\label{equation_momentum_space_hamiltonian}
\end{eqnarray}
HBZ stands for `half of the Brillouin zone', and $\Sigma$ is the sum of all constant terms, $	\Sigma=\frac{N}{2}\kappa(-\vec{m}_A\cdotp\vec{m}_B+3uw+\vec{f}\cdotp\vec{g})$. The momentum-space Hamiltonian $H_{\vec{q}}$ leads to eight bands and the negative energies contribute to the ground state energy. It is written in the basis of $\ket{\Psi}$ as,
\begin{eqnarray}
-i~H_{\vec{q}}=
\begin{bmatrix}
\wp_{11}&\wp_{12}\\
\wp_{21}&\wp_{22}
\end{bmatrix}.\label{equation_momentum_space_hamiltonian_8band_model}
\end{eqnarray}
We use some short hand notations to write these parts more conveniently: $e^{i\hat{q}_1\cdotp\vec{q}}$ = $e^{iq_1}$, $e^{i\hat{q}_2\cdotp\vec{q}}$ = $e^{iq_2}$, $f_x e^{iq_1}$ + $f_y e^{iq_2}$ + $f_z$ = $\tilde{f}$, $g_x e^{iq_1}$ + $g_y e^{iq_2}$ + $g_z$ = $\tilde{g}/3$ = $\tilde{g}_0$, $u(e^{iq_1}+e^{iq_2}+1)$ = $\tilde{u}$, $w(e^{iq_1}+e^{iq_2}+1)$ = $\tilde{w}$. We have assumed $u^\alpha_\gamma=u$ and $w^\alpha_\gamma=w$ for all $\alpha,\gamma=x,y,z$, as mentioned previously.
With these notations, we have,
\begin{eqnarray}\centering
\wp_{11}&=&
\begin{bmatrix}
0 & \lambda_A^z & -\lambda_A^y &
-\begin{pmatrix}
- \kappa_x m_B^x\\+h_x	\\+\lambda_A^x
\end{pmatrix}\\
-\lambda_A^z & 0 & \lambda_A^x & 
-\begin{pmatrix}
- \kappa_y m_B^y\\+h_y \\	+\lambda_A^y
\end{pmatrix}\\
\lambda_A^y & -\lambda_A^x & 0 &
-\begin{pmatrix}
- \kappa_z m_B^z\\+h_z	\\+\lambda_A^z
\end{pmatrix} \\
\begin{pmatrix}
- \kappa_x m_B^x\\+h_x	\\+\lambda_A^x
\end{pmatrix} & 
\begin{pmatrix}
- \kappa_y m_B^y\\+h_y \\	+\lambda_A^y
\end{pmatrix}& 
\begin{pmatrix}
- \kappa_z m_B^z\\+h_z	\\+\lambda_A^z
\end{pmatrix} & 0
\end{bmatrix},\nonumber
\end{eqnarray}
\begin{eqnarray}\centering
\wp_{12}&=&
\begin{bmatrix}
-\kappa_x f_x e^{iq_1} & 0 & 0 & -\kappa_x w_x e^{iq_1} \\
0 & -\kappa_y f_y e^{iq_2} & 0 & -\kappa_y w_y e^{iq_2} \\
0 & 0 & -\kappa_z f_z & -\kappa_z w_z \\
\kappa_x u_x e^{iq_1} & \kappa_y u_y e^{iq_2} & \kappa_z u_z & \begin{pmatrix}
-\kappa_x g_x e^{iq_1}\\-\kappa_y g_y e^{iq_2}\\-k_z g_z
\end{pmatrix} \\
\end{bmatrix},\nonumber
\end{eqnarray}
\begin{eqnarray}\centering
\wp_{22}&=&
\begin{bmatrix}
0 & \lambda_B^z & -\lambda_B^y &
-\begin{pmatrix}
- \kappa_x m_A^x\\+h_x	\\+\lambda_B^x
\end{pmatrix}\\
-\lambda_B^z & 0 & \lambda_B^x & 
-\begin{pmatrix}
- \kappa_z m_A^y\\+h_y \\	+\lambda_B^y
\end{pmatrix}\\
\lambda_B^y & -\lambda_B^x & 0 &
-\begin{pmatrix}
- \kappa_z m_A^z\\+h_z\\	+\lambda_B^z
\end{pmatrix} \\
\begin{pmatrix}
- \kappa_x m_A^x\\+h_x\\	+\lambda_B^x
\end{pmatrix} & 
\begin{pmatrix}
- \kappa_y m_A^y\\+h_y 	\\+\lambda_B^y
\end{pmatrix}& 
\begin{pmatrix}
- \kappa_z m_A^z\\+h_z	\\+\lambda_B^z
\end{pmatrix} & 0
\end{bmatrix}.\nonumber
\end{eqnarray}
We have explicitly written $\lambda_{A/B}^\alpha$, as we have taken them as independent MF variables in some of the trial ansätze, before we were convinced to take a single $\lambda$. Results obtained for MF scheme with uniform $\lambda$, are mentioned as final results. $u_\alpha$ and $w_\alpha$ are also written as vector components, whereas, in our calculations, we have taken $u_\alpha=u$, and $w_\alpha=w$, $\forall\alpha$.
\subsubsection{Self-consistent equations}
Considering half-filling of the bands, only the negative half will contribute to the ground state energy. So, the ground state energy density per unit cell is,
\begin{eqnarray}
E_g=&&\frac{1}{N/2}\sum_{n_1,n_2\in {\rm HBZ}} \sum_{l=0}^3E^{(l)}_{n_1,n_2}\nonumber\\
&&+\sum_{\alpha=x,y,z}\kappa_\alpha (-m_A^\alpha m_B^\alpha+f_\alpha g_\alpha)\nonumber\\
&&+(\kappa_x+\kappa_y+\kappa_z)uw ,\label{eq_E_g}\label{equation_ground_state_energy}
\end{eqnarray}
where $E^{(0,1,2,3)}_{n_1,n_2}$ are the four negative energy bands of the spectrum of $H_{\vec{q}}$ at $\vec{q}=(\frac{n_1}{N_1}2\pi,\frac{n_2}{N_2}2\pi)$. $N_1,N_2$ are number of points in momentum space along $\hat{q}_1,\hat{q}_2$. Minimizing $E_g$ over all the MF parameters, the obtained self-consistent equations are solved using Newton-Raphson method. Denoting the momentum dependent part of Eq(\ref{eq_E_g}) as ${\rm E}$, we define,
\begin{eqnarray}
m_B^\alpha-\frac{1}{\kappa_\alpha}\frac{ \partial}{\partial m_A^\alpha}{\rm E} &=& F_A^\alpha(\vec{X}),\nonumber\\
m_A^\alpha-\frac{1}{\kappa_\alpha}\frac{ \partial}{\partial m_B^\alpha}{\rm E} &=& F_B^\alpha(\vec{X}),\nonumber\\
g_\alpha+\frac{1}{\kappa_\alpha}\frac{ \partial}{\partial f_\alpha^{}}{\rm E} &=& F_f^\alpha(\vec{X}),\nonumber\\
f_\alpha+\frac{1}{\kappa_\alpha}\frac{ \partial}{\partial g_\alpha^{}}{\rm E} &=& F_g^\alpha(\vec{X}),\nonumber\\
w+\frac{1}{(\kappa_x+\kappa_y+\kappa_z)}\frac{ \partial}{\partial u}{\rm E} &=& F_u(\vec{X}),\nonumber\\
u+\frac{1}{(\kappa_x+\kappa_y+\kappa_z)}\frac{ \partial}{\partial w}{\rm E} &=& F_w(\vec{X}),\nonumber\\
\frac{ \partial}{\partial\lambda}{\rm E} &=& F_\lambda(\vec{X}).\label{equation_self_consistent_equations}
\end{eqnarray}
Here, $\vec{X}$ = ( $m_A^x$, $m_A^y$, $m_A^z$, $m_B^x$, $m_B^y$, $m_B^z$, $f_x$, $f_y$, $f_z$, $g_x$, $g_y$, $g_z$, $u$, $w$, $\lambda$ ) is an ordered set of the 15 MF variables, and we denote $\vec{F}(\vec{X})$ = ( $F_A^x$, $F_A^y$ , $F_A^z$, $F_B^x$, $F_B^y$, $F_B^z$, $F_f^x$, $F_f^y$, $F_f^z$, $F_g^x$, $F_g^y$, $F_g^z$, $F_u$, $F_w$, $F_\lambda$ ) to be another set, treated as a 15-dimensional vector, calculated at the given values of $\vec{X}$. We start with some random initial guesses of parameters $\vec{X}^{(0)}$, and keep updating the values for $n=1,2,3,...$ using the Newton-Raphson formula,
\begin{eqnarray}
\vec{X}^{(n+1)} = \vec{X}^{(n)} - \left({\rm J}[\vec{X}^{(n)},\vec{F}(\vec{X}^{(n)})]\right)^{-1} \cdotp\vec{F}(\vec{X}^{(n)}),\qquad\label{equation_newton_raphson_formula}
\end{eqnarray}
until $|X_i^{(n+1)}-X_i^{n}|<\epsilon,~\forall i\in[1,15]$, where $\epsilon=10^{-4}$ in our calculations (for some other MF decompositions with less number of variables, we have gone up to $\epsilon=10^{-5}$). The Jacobian matrix is defined as $\left({\rm J}_{\rm M}[\vec{X}^{(n)},\vec{F}(\vec{X}^{(n)})]\right)_{j,k}$ = ${\partial F_j^{(n)}}/{\partial X_k^{(n)}}$, for $j,k\in[1,15]$.
\\\\\indent
We take the momentum space summation over $100\times100$ grid points to calculate ${\rm E}$. At each parameter value, we minimize the energy over 10 converged set that are obtained with different random initial guesses for the MF variables. For each initial guess, we take 50 iterations for the Newton-Raphson updation, with up to 300 trials of one iteration loop till convergence is achieved. Using the converged values of the MF parameters, we calculate Chern number to identify topological phases~\cite{chern_fukui}.

%*****************************************************************************%
%********************** SECTION - 3 : RESULTS ********************************%
%*****************************************************************************%

\section{Results}\label{section_results}
First we look into the results obtained in the isotropic case, i.e., for $\kappa_{x,y,z}=1$. In subsection \ref{section_anisotropy_results}, we discuss the effect of anisotropy in the exchange couplings.
\subsection{Mean-field results}\label{section_MFresults}
\begin{figure*}[t]\centering
\includegraphics[width=2\columnwidth, height=!]{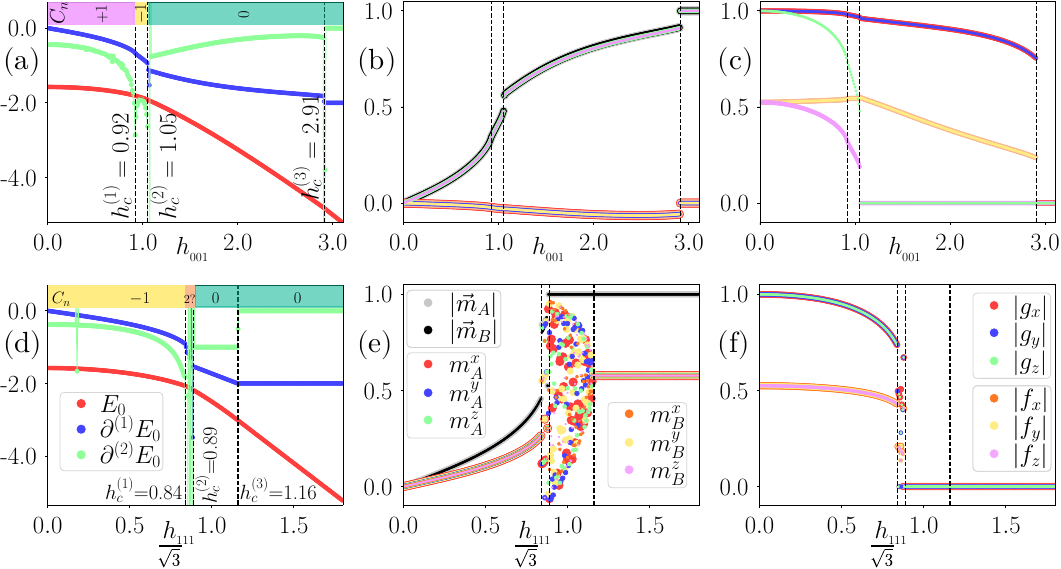}
\caption{Mean-field results for AFM Kitaev model with magnetic field in (a-c) [0,0,1] and (d-f) [1,1,1] direction. (a,d) Ground state energy, and its first and second order derivative with respect to magnitude of the magnetic field. (b,e) Components of magnetization on sub-lattices $A$ and $B$. (c,f) Gauge and matter fields' density. The emergent fields $f_\gamma$ and $g_\gamma$ under this MF scheme behave as conjugate to each other. So, for example in the pure AFM Kitaev limit, if they take simultaneously positive or negative values, both provide the same valid solution. Because in this limit we have, $\sigma^z_j\sigma^z_k\less0$ $\implies$ $\sigma^z_j\sigma^z_k$ = $\la ib_j^zc_j~ib_k^zc_k\ra$ = $-\la ib_j^zb_k^z~ic_jc_k\ra=-gf\less0$ $\implies$ sign($g$) sign($f$)$>0$. That's why we have plotted absolute value of $f_\gamma$ and $g_\gamma$. Chern number $C_n$ is mentioned inside the colored top bar in (a,d). Legends for (a,d), (b,e) and (c,f) are indicated inside the legend box of (d), (e), and (f), respectively.}\label{figure_mf_results}
\end{figure*}
\subsubsection{Magnetic field along [0,0,1] direction}
At $h=0$, ground state (GS) energy per unit cell is $E_g=-1.5746$, an exact match with original result~\cite{kitaev_2006}. Increasing $h$ gradually decreases the $E_g$, eventually makes it linearly varying $E_g\sim-2h$ (two sub-lattice points per unit cell) in large $h$ limit. For $h_z$[FIG.\ref{figure_mf_results}a], system undergoes three phase transitions. At $h_c^{_{(1)}}=0.92$, second order phase transition occurs where system's non-trivial topology persists across a gap closing at $h_c^{_{(1)}}$, and Chern number $C_n$ changes from $+1$ to $-1$. At $h_c^{_{(2)}}=1.05$, a first order phase transition makes the system topologically trivial with $C_n=0$. At $h_c^{_{(3)}}=2.91$, the system enters into SP state via 1st order transition with no change in topology. $m^{z}_{A/B}$ gradually increases with $h_z$, before a discontinuous jump at $h_c^{_{(2)}}$[FIG.\ref{figure_mf_results}b]. (Dis/)continuity in order parameters at critical fields are as per the order of phase transitions. $m^{x,y}_{^{A/B}}$ take almost zero but negative values due to AFM coupling, and beyond $h_c^{_{(3)}}$, they completely vanish.
\\\\\indent
At $h=0$, emerging fields $|g_{\alpha}|\approxeq1$, for all $\alpha=x,y,z$, due to their $Z_2$-structure [FIG.~\ref{figure_mf_results}c]. In absence of magnetic field, the system is integrable and the flux operator expectation can be calculated as $\la \hat{B}_p\ra$ = $g_x^2g_y^2g_z^2$ = $1.000007$, which is in very good agreement with the actual value of unity. For the matter fields, we have got $|f_{\alpha}|=0.52$, for all ${\alpha}=x,y,z$, which is same as the original value~\cite{saptarshi_2007}. Increase in $h_z$ immediately destabilizes $g_z$, but the other fields $g_{x,y}$ are robust up to $h_c^{_{(2)}}$. This is in partial agreement with our previous work with small Kitaev clusters~\cite{kitaev_cluster_dynamical_pervez}. At $h_c^{_{(2)}}$, $g_z$ is completely destroyed, whereas $g_{x,y}$ pick up dynamics, and that is reflected through the decay in their density. At $h_c^{_{(3)}}$, these two fields vanish too, indicating the onset of SP phase.
\\\\\indent
The work by Nasu et al.~\cite{sucessive_majorana_KSL_001_nasu} captures the first two critical fields at $\sim0.834$ and $1.006$ (we have multiplied by 2, as they considered $\vec{S}_j=\vec{\sigma}_j/2$, instead of just the Pauli operators), which are very close to our data, while the third critical field is absent in their calculations. With ED, they identify a single transition at $h_z\sim0.61$, which is exactly reproduced in our calculations too~[\ref{section_EDresults}].
\subsubsection{Magnetic field along [1,1,1] direction}
Now we focus at the results with a conical field. With $h_{111}$[FIG.\ref{figure_mf_results}d], two 1st order and one 2nd order phase transition occurs at critical fields $h_c^{_{(1)}}=0.84\sqrt{3}$, $h_c^{_{(2)}}=0.89\sqrt{3}$, and $h_c^{_{(3)}}=1.16\sqrt{3}$. As soon as the $h_{111}$ becomes non-zero, it opens up a gap in the energy spectra, and the resulting phase is topologically non-trivial with $C_n=-1$. Survival of the emergent structures till $h_c^{(1)}$ indicates partially robust QSL. Beyond that, it enters into an intermediate region, whose exact characteristics is still under debate.
\\\\\indent
The isotropic Kitaev model with a conical magnetic field has cubic symmetry, and thus the Cartesian components of magnetization should coincide with each other. This is indeed true in low-field QSL and high-field SP phase, as we observe in FIG.~\ref{figure_mf_results}e. But in the intermediate phase, something peculiar happens. In between $h_c^{_{(1)}}$ and $h_c^{_{(2)}}$, the gauge and matter fields fluctuate but are not destroyed completely [FIG.\ref{figure_mf_results}f]. So, the liquidity persists partially. When we calculate the Chern number, it is ill-defined (fluctuates between 0 and 2~\cite{ill_defined_chern_number}), as already reported in Ref.~\cite{emergent_magnetic_order_KSL_111}. The magnetization components loose the cubic symmetry and converge to random values. The magnitude $|\vec{m}_{A/B}|$ also fluctuate, and stays below 1, indicating strong quantum effects. All these make us believe that it is a disordered phase with randomly directed magnetization, and poorly defined topology.
\\\\\indent
In our MF calculation, we identify a second intermediate phase. In between $h_c^{_{(2)}}$ and $h_c^{_{(3)}}$, $m^{x,y,z}_{^{A/B}}$ converge to arbitrary values, but, they do follow the constraint of $(m_{^{A/B}}^x)^2$ $+(m_{^{A/B}}^y)^2$ $+(m_{^{A/B}}^z)^2=1$~[FIG.~\ref{figure_mf_results}e]. The emergent structures are almost vanished in this phase, and we calculate $C_n=0$. The magnetization operator is behaving like a classical vector (with the full unit magnitude) which points along any direction, indicating the presence of large classical GS degeneracy. So we paint the whole picture in this region as follows. At $h_c^{_{(2)}}$, the system already enters into a classical regime. The magnetization vector actually precesses around $\hat{h}_{111}$ by an angle $\theta$. As we increase $h_{111}$, $\theta$ gradually decreases, and at $h_c^{_{(3)}}$, it becomes zero, indicating the commencement of SP phase where the spins are aligned along the direction of the magnetic field.
\\\\\indent
In between $h_c^{_{(2)}}$ and $h_c^{_{(3)}}$, presence of large GS degeneracy is in accordance with previous works that show large number of low lying eigen-states exist in this intermediate region~\cite{u1_KSL_natcom,magnetic_field_driven_qpt_nandini,robust_non_abelian_gapless_prb}. Under the present MF framework, this phase in not a spin liquid, as the emergent structures are (almost) vanished. All MF work in literature, assume the cubic symmetry in the observables, to keep the number of MF variables less, so that the convergence is achievable. As a consequence, this rich structure remained hidden. Our work goes beyond this assumption, considers the variables to be direction-dependent, and thus able to bring out the constrained structure.
\\\\\indent
In summary, as we increase $h_{111}$, the emergent fields remain close to their zero-field values indicating persisting QSL, starts to fluctuate at $h_c^{_{(1)}}$ indicating commencement of a disordered phase, vanish beyond $h_c^{_{(2)}}$ depicting the absence of QSL and emergence of classical degenerate ground states, and remain zero across $h_c^{_{(3)}}$ while approaching SP state.
\subsection{Exact-diagonalization results}\label{section_EDresults}
\begin{figure*}\centering
\includegraphics[width=2\columnwidth,height=!]{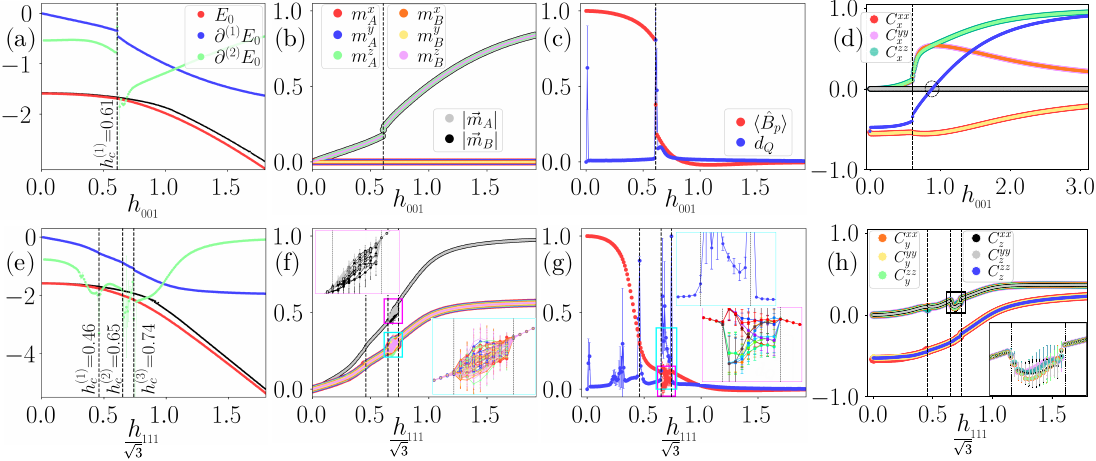}
\caption{Exact diagonalization results for magnetic field in (a-d) [0,0,1], and (e-h) [1,1,1] direction. (a,e) Ground state energy and its derivatives with respect to magnetic field strength. MF energy is shown in black line for comparison. (b,f) Magnetization components (not averaged, and explicitly plotted for all sites). (c,g) Expectation value of the flux operator $\la \hat{B}_p\ra$, and quantum distance $d_Q$. The set of twelve flux operator values are not averaged, and they have been plotted one top of another, to explicitly show the fluctuation in between $h_c^{(2)}$ and $h_c^{(3)}$. (d,h) Spin-spin correlations on the bonds: 21-22 ($z$-type), 22-23 ($x$-type), and 23-24 ($y$-type). Legends for (a,e), (b,f) and (c,g) are indicated inside the legend box of (a), (b), and (c), respectively. Correlations with different colors are split into multiple legend boxes, and are kept in (d) and (h).}\label{figure_ed_results}
\end{figure*}
Under MF approximation, some of the correlations are always left out, and it is important to validate MF results by some other method. Here we choose to study the system with ED using Lanczos algorithm. We take average over the results obtained from different initial random configurations. Standard deviations (SD) are shown as vertical bars in all panels of FIG.\ref{figure_ed_results}.
\subsubsection{Magnetic field along [0,0,1] direction}
GS ED energy, $E_{\rm GS}=-1.5881$ per unit cell, is very close to MF data. The resemblance between MF and ED results for all $h$ values show that we have obtained GS via MF, very close to the exact one. This is possible only because we have dealt with large number of MF variables, that describes the system more accurately. Increase in $h$ makes $E_{\rm GS}$ to decrease, and eventually to vary as $\sim -2h$ in large $h$ limit. Due to finite size effects, ED captures only a 1st order phase transition at $h_c^{_{(1)}}=0.61$, and other critical points are absent [FIG.~\ref{figure_ed_results}a].
\\\\\indent
Absence of non-Kitaev interaction makes $A$ and $B$ sub-lattice to experience the magnetic field equally [FIG.\ref{figure_ed_results}b]. As $h$ is along $\hat{z}$-direction, $m^{{x,y}}_{^{A/B}}$ remain zero throughout the zero-field QSL phase to the high-field SP state. At $h_c^{_{(1)}}$, $m^z_{^{A/B}}$ changes discontinuously because of the 1st order transition, and asymptotically approaches unity at large $h$ limit.
\\\\\indent
Flux operator $\hat{B}_p$ is defined as the product of pure-Kitaev Hamiltonian terms around a plaquette, that effectively reduces to product of the Pauli operators projected along the outward directions from each site on a plaquette (one of them, $\hat{B}_p=\sigma_{19}^{y}$$\sigma_{20}^{z}$$\sigma_{21}^{x}$$\sigma_{22}^{y}$$\sigma_{23}^{z}$$\sigma_{24}^{x}$, is shown in FIG.\ref{figure_schematics}b). At $h_z=0$, $\la \hat{B}_p\ra$ starts from $+1$, monotonously decreases with increasing $h_z$, shows a discontinuous jumps to very low values at the critical field. At large $h_z$ values, $\la \hat{B}_p \ra$ completely vanishes, indicating equal mixing of the $Z_2$ vortices; thus QSL is completely destroyed.
\\\\\indent
To calculate the gradual fidelity, we define quantum distance $d_Q^2(h)=1-|\braket{\psi_{h}}{\psi_{h+\Delta h}}|^2$~\cite{Resta2011}, where $\ket{\psi_h}$ is the ED GS at $h$. $d_Q\sim0$ indicates that the two consecutive GS are very similar (`close' to each other). Whereas, a large $d_Q$ tells that consecutive GS are independent of each other. When we introduce non-zero $h_z$, the GS is very different from what it was in pure-Kitaev limit, and we get large $d_Q$ [FIG.~\ref{figure_ed_results}c]. GS evolves adiabatically with only a rapid transition at the critical field. Beyond $h_c$, GS smoothly connects to the SP state at large $h$ limit.
\\\\\indent
Zero temperature spin-spin correlation is defined as $C^{\alpha\beta}_{jk}= \expval{\sigma^{\alpha}_{j} \sigma^{\beta}_{k}}{\rm GS}$. In pure-Kitaev limit, the exact value of this correlation comes from the matter sector only, as $C^{\alpha\alpha}_\gamma=-\la ic_jc_k \ra~\delta_{\alpha,\gamma}$, and all the cross correlations are zero~\cite{saptarshi_2007}. At zero-field, we get the non-zero correlations as $C^{xx}_{x}$ = $C^{yy}_{y}$ = $C^{zz}_{z}$ = -0.529367 with SD of 0.031519; actual value $-0.524866$ is well within the error bar. $h_z$ does not build $x-x$ and $y-y$ correlation in the $z$-type bonds, and we get $C^{xx}_{z}=C^{yy}_{z}=0~\forall h_z$. $C^{yy}_{x}$ and $C^{xx}_{y}$ undergoes a subtle discontinuity at $h_c=0.61$, supporting the 1st order phase transition. As $h_z$ starts to align the spins along $\hat{z}$, $C^{yy}_{x}=C^{xx}_{y}\rightarrow0$ in the limit $h_z\rightarrow\infty$. $C^{zz}_{x}=C^{zz}_{y}$ starts from zero at $h_z=0$, jumps up at $h_c^{_{(1)}}$, and gradually saturates to unity in large $h_z$ limit. Interestingly, on a $z$-type bond, all $x-x,y-y,z-z$ correlations exactly vanish at $0.88<h_{0}<0.89$ [FIG.\ref{figure_ed_results}d, black dotted circle]. So, at $h_{0}$, the system effectively reduces to decoupled $x$-$y$ chains, which is in accordance with previous work of dimensional reduction in KSL~\cite{dimensional_reduction_KSL}. Using ED GS, we have calculated the distanced neighbor correlations at $h_{0}=0.8864$. We get $|C^{yy}_{j,j+k}|$ $\sim$ $|C^{xx}_{j,j+k}|$ = 0.536361, 0.265968, 0, 0, for $k=1,2,3,4$ respectively. Whereas $C^{zz}_{j,j+k}$ = 0.537345, 0.174880, 0.195425, 0.177797 for $k=1,2,3,4$ respectively. So, $x-x,y-y$ correlations are short ranged, as they survive only up to second nearest neighbor. Whereas, $z-z$ correlations saturate at some finite value for long-distance neighbors, and that produce the magnetization in $\hat{z}$-direction as in FIG.\ref{figure_ed_results}b.
\subsubsection{Magnetic field along [1,1,1] direction}
In this subsection, we showcase the ED results with $h_{111}$. Despite finite size effects, ED captures all three critical transitions with quantitatively different (than MF) fields, $h_c^{_{(1)}}=0.46\sqrt{3}$, $h_c^{_{(2)}}=0.65\sqrt{3}$, $h_c^{_{(3)}}=0.74\sqrt{3}$, and all of the transitions are identified as 2nd order [FIG.~\ref{figure_ed_results}e]. All of the flux operators are destabilized equally in the low field partially robust QSL, and in high field SP phase. But, in between $h_c^{_{(2)}}$ and $h_c^{_{(3)}}$, $\la \hat{B}_p\ra$ fluctuates rapidly as captured by its large SD (FIG.\ref{figure_ed_results}g, magenta {\color{magenta}$\Box$} inset) calculated from different Lanczos initializations. In this region, ED magnetization magnitude and the Cartesian components fluctuate rapidly with high SD (FIG.\ref{figure_ed_results}f, magenta {\color{magenta}$\Box$} inset). Apart from large SD, we observe substantial non-zero $d_Q$ in this region, indicating that consecutive GS are independent from each other. The non-zero correlations at zero-field, gradually increases with increase in $h$ at equal rate, crosses zero, and eventually saturates to $1/3$. The other correlations also grow from zero and asymptotically approaches $1/3$. In between $h_c^{_{(2)}}$ and $h_c^{_{(3)}}$, the correlations become weak and they fluctuate with large SD, as depicted in black {\color{black}$\Box$} inset box of FIG.\ref{figure_ed_results}h.
\\\\\indent
It is worth mentioning that in the model under consideration, a triple (quasi-)degeneracy in the ground state is expected owing to the Ising topological quantum field theory~\cite{u1_KSL_natcom}. We observe such (quasi) degeneracy to occur till $h_c^{(3)}$ [FIG.~\ref{figure_excited_lanczos_states}b]. Interestingly, higher $d_Q$ value in this region means the GS at each field strength changes rapidly having very small fidelity between them, which is suggested to be a spin-glass phase for Kitaev model on a ribbon geometry using density matrix renormalization group technique~\cite{emergent_glassiness_yogendra}. In our ED calculations, near $h_{111}\sim0.23\sqrt{3}$, the lowest three Lanczos states are degenerate within $\Delta E\sim0.0002$; but that affects the cubic symmetry in the magnetization components minimally. Whereas, in between $h_c^{(2)}$ and $h_c^{(3)}$, they stay within $ E_0+0.002$, but the results appear to suffer a numerical convergence of the GS to random directions in a degenerate subspace. This hints us, the GS in this phase might be fundamentally different. The sudden drop in the correlation also indicate the system possibly orients the spins in specific directions depending on the bond interaction, and the large SD is a signature of the different accessible configurations of that orientation. Comparing with the MF analysis presented in~\ref{section_MFresults}, we believe the ED result is also a indicator of the classical degenerate GS, with some finite size effects (for example, fluctuations in the magnetization become minimal, unlike MF). Interestingly, we do find similar fluctuations in topological entanglement entropy~\cite{magnetic_field_driven_qpt_nandini}, and in genuine multipartite negativity for intermediate magnetic field~\cite{lyu2025multipartyentanglementloopsquantum}.
\begin{figure}[h]\centering
\includegraphics[width=1\columnwidth,height=!]{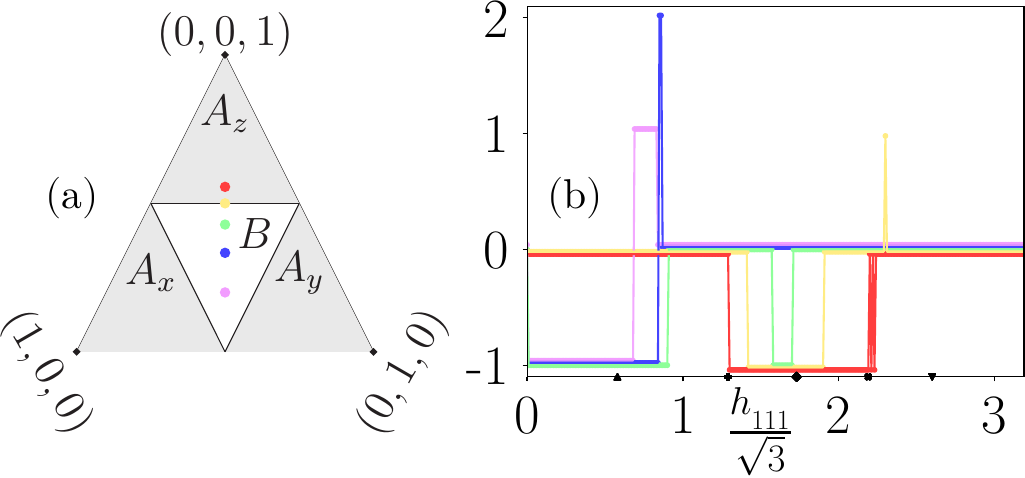}
\caption{(a) Gapless ($B$) and gapped ($A_{x,y,z}$) phases (in absence of magnetic field) in the Kitaev coupling parameter space on the triangle having $\kappa_x+\kappa_y+\kappa_z=1$. The chosen coupling constants when projected on this plane, are indicated by colored dots. (b) Chern number $C_n$ as we vary strength of the magnetic field. The color coding indicates the coupling strengths as indicated on the Kitaev's parameter triangle in the left panel. We have given a slight shift along the vertical axis for visual clarity. The black dots on the field axis indicate the $h_{111}$ values for which the band spectra is shown in FIG.~\ref{figure_anisotropic_kitaev_energy_band}, for the parameter values $\kappa_x=\kappa_y=1$, $\kappa_z=2.5$ (red line here).}
\label{figure_anisotropic_kitaev_topology}
\end{figure}
\begin{figure*}\centering
\includegraphics[width=2\columnwidth,height=!]{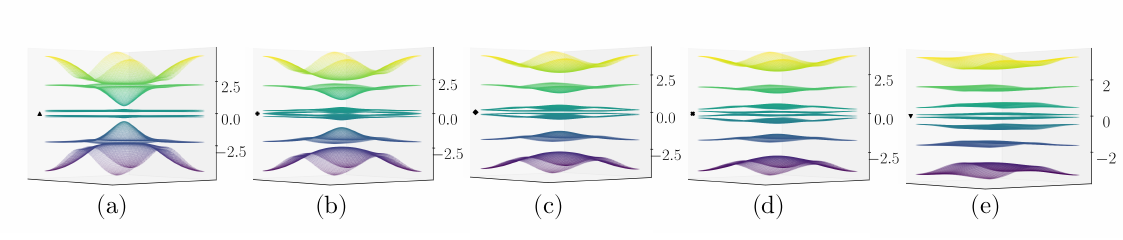}
\caption{The eight band energy spectra in momentum space for $\kappa_x=\kappa_y=1$, $\kappa_z=2.5$ (corresponds to red line in FIG.~\ref{figure_anisotropic_kitaev_topology}). Various strengths of the magnetic field are $h_{111}/\sqrt{3}$ = (a) 0.58, (b) 1.29, (c) 1.73, (d) 2.19, (e) 2.60, indicated by different black dots on the horizontal axis of FIG.~\ref{figure_anisotropic_kitaev_topology}b. The system makes topological transition at (b) and (d).
}
\label{figure_anisotropic_kitaev_energy_band}
\end{figure*}
\subsection{Anisotropic Kitaev couplings}\label{section_anisotropy_results}
Kitaev model is known to host both gapless and gapped energy spectra depending on the relative strength of the exchange couplings. When they obey the triangular inequality ($|\kappa_\alpha|+|\kappa_\beta|\geq|\kappa_\gamma|$), the spectrum is gapless. Resulting QSL is actually a topologically non-trivial one, which can be calculated by opening a gap in the spectra by introducing a perturbation. When the $\kappa_\alpha$'s are beyond the triangular inequality, the spectrum is gapped, and it is topologically trivial. Here we study the fate of this gapped trivial QSL when subjected to a magnetic field.
\\\\\indent
To this end, we investigate the effect of anisotropy in the Kitaev couplings $\kappa_{x,y,z}$ in presence of magnetic field in the $[1,1,1]$ direction. In absence of $h_{111}$, the Kitaev phase diagram is shown in FIG.~\ref{figure_anisotropic_kitaev_topology}a. We choose the coupling strength values as $\kappa_x=\kappa_y=1$, and $\kappa_z=0.5,1,1.5,2,2.5$. As soon as a perturbative magnetic field is turned on, the gapless phase (`$B$' in FIG.~\ref{figure_anisotropic_kitaev_topology}a) acquires a gap and become topologically non-trivial with $C_n=-1$. For $\kappa_z=0.5$, the system makes a topological transition into $C_n=+1$ at a certain field value, and eventually makes another transition to the trivial phase when the external field is increased further. For $\kappa_z=1.5$, the system goes from $-1\rightarrow0\rightarrow-1\rightarrow0$ at certain strengths of $h_{111}$.
\\\\\indent
The particularly interesting case is when the system is otherwise topologically non-trivial ($\kappa_x=\kappa_y=1$, $\kappa_z=2.5$). As we observe in FIG.~\ref{figure_anisotropic_kitaev_topology}b, in presence of the conical magnetic field's intermediate strength, topology shows a reentrant behavior. We can understand this in the following way. The external field suppresses the anisotropic AFM interactions (stronger effect on the larger $\kappa_\alpha$), effectively bringing them back within the triangular inequality, and thus making a topological transition. Our numerical investigation suggests that the effective couplings can even be written in an approximate quantitative way as $\kappa_\alpha^{\rm eff}=\kappa_\alpha/|g_\alpha|$, for $\alpha=x,~y,~z$. The $\kappa_\alpha^{\rm eff}$'s when projected to a triangular parameter space (similar to FIG.~\ref{figure_anisotropic_kitaev_topology}a), the topologically non-trivial phase appears when the effective couplings approximately belong to the inner triangle. We demonstrate this correspondence between change in topology and $\kappa_\alpha^{\rm eff}$, in~\cite{effective_couplings_with_anisotropy}. With increase in the field strength, the system eventually becomes spin polarized, thus making a transition to topologically trivial phase. The topological region is quite robust for a wide range of the intermediate field strength. We also plot the energy band in FIG.~\ref{figure_anisotropic_kitaev_energy_band} to explicitly demonstrate the gap closure during the topological transitions.

%*****************************************************************************%
%************************** SECTION - 4 : CONCLUSION *************************%
%*****************************************************************************%

\section{Conclusion}\label{section_conclusion}
We have performed a Majorana mean-field analysis with an extensive number of variables to bring out the rich physics of the Kitaev-Zeeman model. In contrast to previous mean-field works in the literature~\cite{thaigarajan_topological_transition,majorana_mean_field_kruger,tuning_topological_order_PRL,novel_chiral_QSL_kitaev_dm,intermediate_gapless_phase,sucessive_majorana_KSL_001_nasu}, we consider the variables to be direction-dependent. The novelty in our work lies in the successful convergence of the self consistent equations, despite considering so many variables.
\\\\\indent
When the Zeeman field is along a Cartesian axis, we find that the emergent fields are stabilized depending on their directionality (on what type of bond they are initially defined). Interestingly, the stabilization of these emergent structures is an direct indication that the QSL remains partially robust up to certain critical field, as established in the literature by means of different numerical methods~\cite{shifeng_anyon_dynamics,feng2026magneticfieldinducedphenomena}. Upon application of a conical magnetic field, we show the presence of two phases in the intermediate strength of the field. The first one being magnetically disordered, where the emergent fields converge to random values, and the Chern number becomes ill-defined. In the next intermediate phase, the magnetization vectors behave as classical ones, with their direction directed along arbitrary directions. With increase in the external field strength, the magnetization vectors, gradually becomes more aligned to the magnetic field's direction, and after a critical strength of the Zeeman field, enters the trivial spin polarized phase. The presence of two intermediate phases has been explored using cluster mean-field calculations~\cite{emergent_magnetic_order_KSL_111}, but the existence of degenerate classical ground states is not shown. Performing exact diagonalization calculations, we qualitatively re-establish some of the conclusions from our mean-field analysis.
\\\\
Finally, we investigate the effect of anisotropy in the Kitaev couplings in the presence of a conical magnetic field, and find that topology shows a reentrant behavior in the `gapped' phase (of the unperturbed Kitaev model) at some intermediate strength of the external field. Our numerics suggest that the Kitaev couplings are effectively rescaled, which can even be written in an approximate quantitative equation, and the reentrance of topology occurs when these effective couplings satisfy the triangular inequality. It is intriguing to extend our study in presence of other non-Kitaev interactions, and to check if in a more realistic model (including Heisenberg, $\Gamma$, $\Gamma'$ terms), a proximate topologically non-trivial QSL survives, that would in principle manifest its signature through a half-quantized thermal Hall response.
\\\\\indent
It is worthwhile to make further similar investigations for the Kitaev model in higher dimensions~\cite{naveen_2008,defect_production_quench_dynamics}. It will also be interesting if we find any correspondence under the mean-field scheme to the emerging narrow bands in the eigen-spectra when a competing Heisenberg interaction (opposing the Kitaev coupling) is present~\cite{pervez2026QOE}. It is compelling to investigate the fate of recently proposed emergent quadrupolar order~\cite{monodip_quadrupolar} in such a system, and what is its signature under a mean-field scheme. All these questions can be addressed extending the present work, and we leave them as future scope.

%*****************************************************************************%
%***************** SECTION - DATA AVAILABILITY STATEMENT *********************%
%*****************************************************************************%

\section*{Data Availability Statement}
The data supporting this study's findings are available upon reasonable request from the corresponding author.

%******************************************************************************%
%************************ SECTION - ACKNOWLEDGEMENT ***************************%
%******************************************************************************%

\section*{Acknowledgement}
The author gratefully acknowledges Saptarshi Mandal for suggesting the problem, for insightful discussions, and for carefully reading the manuscript. Numerical calculations were partially carried out at SAMKHYA (High-Performance Computing facility), provided by the Institute of Physics, Bhubaneswar.

%******************************************************************************%
%*************************** SECTION - BIBLIOGRAPHY ***************************%
%******************************************************************************%

\bibliography{bibfile.bib}

%******************************************************************************%
%****************************** SECTION - APPEXNDIX ***************************%
%******************************************************************************%

\appendix
\section{Appendix}
\subsection{Benchmark test of Lanczos algorithm}\label{section_appendix_lanczos_benchmark}
\begin{figure}[h]\centering
\includegraphics[width=\columnwidth,height=!]{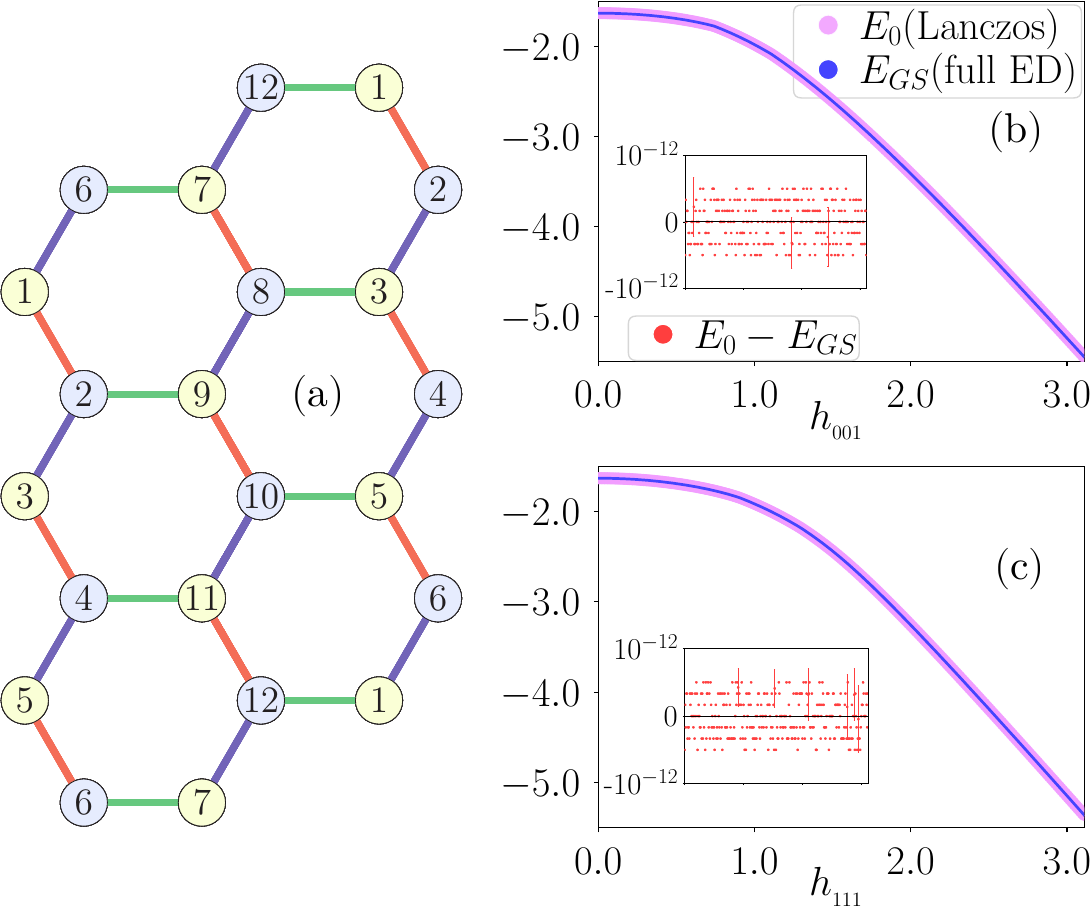}
\caption{Benchmark test for the Lancos ground state energy $E_0$ with respect to the actual ground state energy $E_{GS}$. (a) The 12-site cluster that we have used for benchmarking, and the magnetic field is in (b) [0,0,1]-direction, and (c) [1,1,1]-direction. In the insets, we have plotted difference between the two energies.}\label{figure_lanczos_benchmark}
\end{figure}
As we have already observed [FIG.\ref{figure_ed_results}a,e], the MF ground state is very close to the Lanczos ground state, and thus it passes the benchmark test. But, what about the Lanczos state itself? Here we do a benchmark test for the Lanczos, while comparing the energy obtained for a 12-site Kitaev cluster using full-diagonalization. Throughout this paper, for Lanczos algorithm, we have set the Krylov space dimension at 200. This is good enough to give the ground state energy, as we will see in the convergence test. We have used a 12-site Kitaev cluster with periodic boundary condition, as depicted in FIG.\ref{figure_lanczos_benchmark}a, to execute the full exact diagonalization. For benchmarking, we have used 11 Lanczos initializations. The results for two different orientation of the magnetic field is shown in FIG.\ref{figure_lanczos_benchmark}b,c. They match exactly. To have a closer look at the difference, we plot $\Delta E=E_{\rm Lanczos}-E_{\rm full ~ED}$ in the insets. A curious reader might have the doubt why does $\Delta E$ become negative sometimes. According to the variational principle of quantum mechanics, it should always be positive. To answer that, we want to mention, we have taken precision up to $10^{-12}$ for the energies, and any difference (positive or negative) within this window comes from the machine precision only.
\subsection{Convergence test of Lanczos algorithm}\label{section_appendix_lanczos_convergence}
\begin{figure}[h]\centering
\includegraphics[width=\columnwidth,height=!]{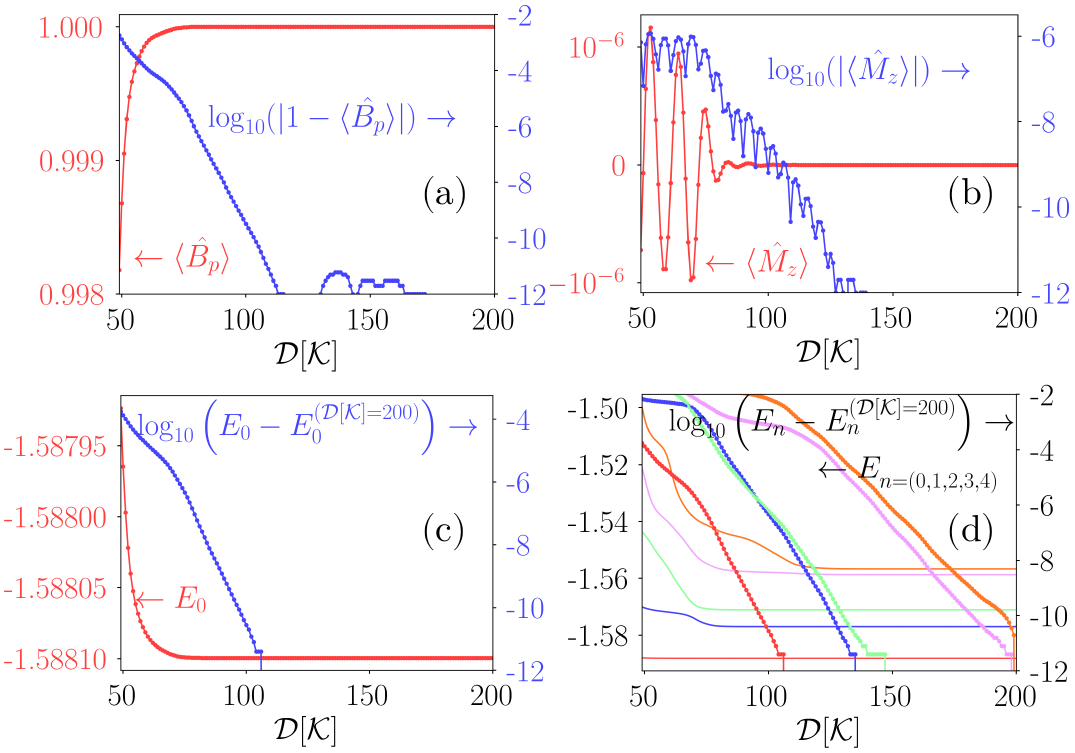}
\caption{Convergence test for Lanczos algorithm on the 24-site lattice under consideration, in the pure Kitaev limit. As we vary the dimension of Krylov space ($\mathcal{D}[\mathcal{K}]$), here we plot the expectation of (a) flux operator, (b) total magnetization along $\hat{z}$-direction, (c) ground state energy, and (d) few excited states' energy, along the left vertical axis. Along the right vertical axis, on a logarithmic scale, we plot deviation of the same quantities from their corresponding values obtained for $\mathcal{D}[\mathcal{K}]=200$. These plots are based on only one realization of the initial random vector, and already shows convergence of the order $10^{-12}$ for $\mathcal{D[K]}\sim120$. }\label{figure_lanczos_convergence}
\end{figure}
In the previous section, the Lanczos algorithm passes the benchmark test. But what should be the dimension of Krylov space ($\mathcal{D[K]}$) to achieve convergence in the 24-site honeycomb lattice under consideration? To find that, we calculate flux operator expectation, total magnetization, ground and some of the excited states' energy, as we vary $\mathcal{D[K]}$. As we observe in FIG.~\ref{figure_lanczos_convergence}, for $\mathcal{D[K]}\sim120$, we already achieve incredible convergence. In practice, we have taken $\mathcal{D[K]}=200$. In addition to that, we take average over different random initializations of the Lanczos process.
\subsection{Lanczos energies of the excited states}\label{section_lanczos_excited_states}
\begin{figure}[H]\centering
\includegraphics[width=\columnwidth,height=!]{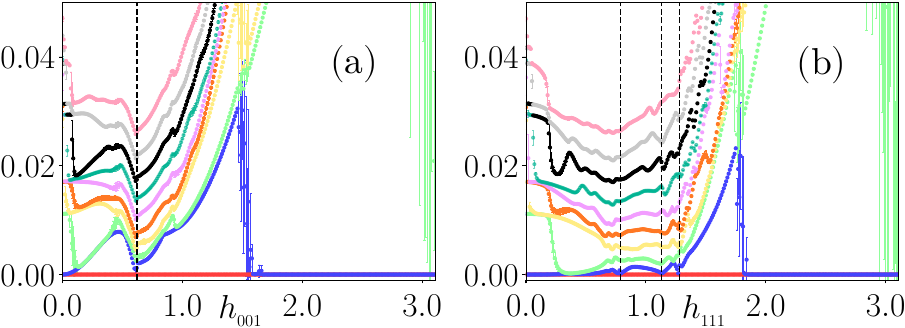}
\caption{Ten lowest energies as calculated using Lanczos algorithm, shifted by the ground state energy, for magnetic field in (a) [0,0,1] and (b) [1,1,1] direction. The black dotted lines indicate the critical field strength as calculated by change in slope of $E_{\rm GS}$.}
\label{figure_excited_lanczos_states}
\end{figure}
\subsection{Plaquettes}\label{section_appendix_plaquettes}
The 24-site honeycomb is defined on a torus with 12 plaquettes. Seven of them are easily identifiable in FIG.\ref{figure_schematics}(b). The other five plaquettes are:\\
\begin{figure}[H]\centering
\includegraphics[width=0.9\columnwidth,height=!]{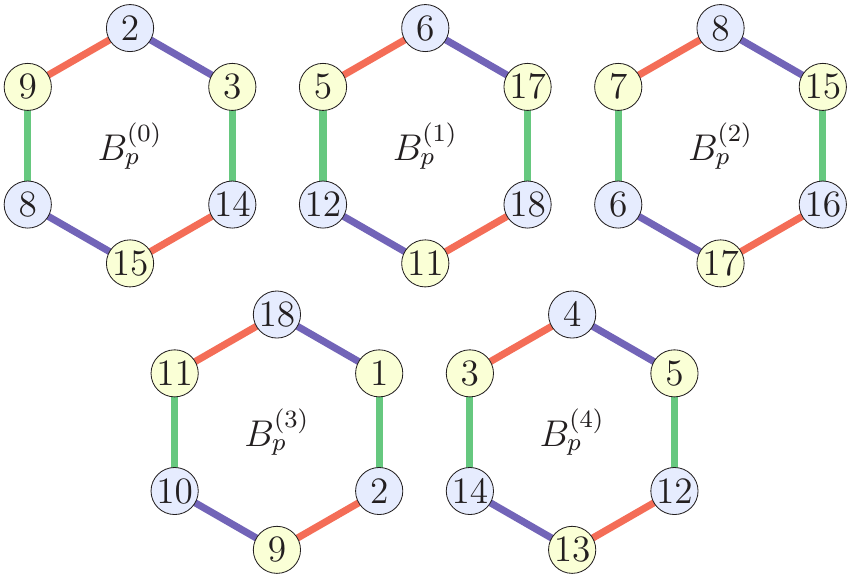}
\caption{Five plaquettes and their corresponding site arrangements in the periodic 24-site cluster under consideration.}\label{figure_plaquettes}
\end{figure}

\end{document}